\documentclass[letterpaper, 12pt]{interact}

\usepackage[doublespacing]{setspace}
\usepackage[numbers,sort&compress]{natbib}
\bibpunct[, ]{[}{]}{,}{n}{,}{,}
\renewcommand\bibfont{\fontsize{10}{12}\selectfont}

\usepackage{color,soul}
\usepackage{hyperref}
\usepackage[document]{ragged2e}

\begin{document}
\title{Automated surface feature selection using SALSA2D: Assessing 
distribution of Elephant carcasses in Etosha National Park}

\author{
\name{L.A.S Scott-Hayward\textsuperscript{a}\thanks{CONTACT L.A.S. Scott-Hayward. Email: lass@st-andrews.ac.uk}, M.L. Mackenzie\textsuperscript{a}, C.G. Walker\textsuperscript{b}, G. Shatumbu\textsuperscript{c}, W. Kilian\textsuperscript{c} and P. du Preez\textsuperscript{d}}
\affil{\textsuperscript{a}School of Mathematics and Statistics, University of St Andrews, KY16 9LZ, Fife, Scotland; \textsuperscript{b}Department of Engineering Science, University of Auckland, 70 Symonds Street, Auckland, New Zealand; \textsuperscript{c}Etosha Ecological Institute, PO Box 6, Okaukuejo via Outjo, Ministry of Environment, Forestry and Tourism, Namibia; \textsuperscript{d}African Wildlife Conservation Trust, PO box 97401, Windhoek, Namibia}
\affil{\textsuperscript{a}ORCiD: LSH (0000-0003-3402-533X), MLM (0000-0002-8505-6585)}
}

\maketitle 

\vspace{1cm}

\section*{Open Research Statement}
The data and code are provided as private-for-peer review but can be made public if accepted for publication.  The files can be found at the github site of the first/corresponding author: \url{https://github.com/lindesaysh/MIKE}. Additionally, the data file will be made public and permanently archived in the St Andrews PURE repository.

\begin{keywords}
CReSS (Complex Region Spatial Smoother); Spatially Adaptive; SALSA; spline; point process; GAM; regression
\end{keywords}

\newpage
\begin{abstract}


This paper describes the development of an automated knot selection method (selecting number and location of knots) for bivariate splines in a pure regression framework (SALSA2D). To demonstrate this approach we use carcass location data from Etosha National Park (ENP), Namibia to assess the spatial distribution of elephant deaths. Elephant mortality is an important component of understanding population dynamics, the overall increase or decline in populations and for disease monitoring. 

The presence only carcass location data were modelled using a downweighted Poisson regression (equivalent to a point-process model) and using developed method, SALSA2D, for knot selection. The result was a more realistic local/clustered intensity surface compared with an existing model averaging approach. 

Using the new algorithm, the carcass location data were modelled using additional environmental covariates (annual rainfall, distance to water and roads). The results showed high carcass intensity close to water holes ($<$3km) and roads ($<$2km) and in areas of the park with average rainfall ($\sim$450mm annually). Some high risk areas were identified particularly in the north east of the park and the risk of death does not always coincide with elephant distribution across the park. These findings are an important component in understanding population dynamics and drivers for population and park management. Particularly for controlling elephant numbers and/or mitigation of anthrax or other disease outbreaks. 


\end{abstract}


\newpage

\section*{Introduction}
Spline based regression is a well established method for estimating relationships when the functional form between an expected response and a set of covariates is unknown and non-linear. Splines are restrictive enough to benefit from parametric estimation and general enough to approximate a wide range of smooth functions. Within the parametric estimation framework, there are two main approaches to estimating these functional forms, a penalised or un-penalised approach.  In the former, a penalty term which includes a smoothing parameter is used to determine the wiggliness of the spline and in the latter approach the appropriate wiggliness is determined by judicious placement and number of knots. While penalised approaches are covered extensively, for example in \citet{wood2017} or \citet{eilers2010}, this paper focuses on methods for judicious placement and number of knots, the un-penalised approach. 

\citet{Walk2010} presented an algorithm for adaptively placing knots called SALSA - Spatially Adaptive Local Smoothing Algorithm. It is an adaptive knot selection approach, with the number and location of the knots being determined by the solution process. The algorithm is not an all possible subsets approach but combines a local-search strategy with a restricted forward/backward regression approach to significantly reduce the number of models evaluated at each iteration. The paper demonstrated SALSA for univariate splines and found it to be an ``intuitive solution that is naturally able to accommodate local changes smoothness``. 

\citet{scott2014} developed the CReSS (Complex Region Spatial Smoother) approach to allow bivariate smooths to cope with complex topographies to respect the natural boundaries encountered by animals, e.g. complex coastlines or lakes. This approach does not choose the number and location of knots for a bivariate spline but achieves spatially adaptive surfaces via a judicious weighting (model averaging) of a variety of candidate surfaces.  Each model uses a set of exponential basis functions located at a fixed number of `space-filled' knots \cite{Johnson1990} and fixed radius to determine the influence of each basis. These models result in surfaces ranging from the very simplistic (via small numbers of knots with basis functions with a relatively global influence) to very complex (via large numbers of knots with basis functions with a relatively localised influence).

This approach (CReSS with model averaging) has been shown to perform well against other model-based alternatives developed for data sets with internal exclusion zones (such as coastlines and island systems) and is finding use in a range of ecological applications \cite{russell2016, dunlop2017, harris2018}.

However, while this approach  has been shown to produce reliable results in many cases \citep{scott2014}, this procedure can be complicated, in terms of model handling and difficult to assess model fit and to provide confidence intervals. A paper by \citet{dormann2018} highlights some of the limitations of a model averaging approach. Namely, the authors show that estimating model weights introduces unknown and unaccounted for uncertainty and that confidence intervals for model-averaged predictions rarely achieve nominal coverage. They also state that model-averaging is most useful when the predictive error of contributing model predictions is dominated by variance (as opposed to bias), and if the covariance between models is low. We argue that ecological data, including the carcass data seen here, is often highly variable with limited covariates and thus could result in prediction errors dominated by variance. Additionally, given the CReSS with model averaging approach averages models with the same covariates but different parameterisations, there is also likely to be high covariance between competing models, rendering the model averaging approach less appropriate.

Further, when the spatial patterns are particularly unusual (e.g.~stripe-like features or local hotspots are genuinely present) we have found that a model-averaging approach can result in overly smooth surfaces which mask these unusual, but important, patterns. This may result for a variety of reasons: under the original CReSS approach the space-filled knots are fixed in position for a given knot number, and the extent that each basis function is local (or global) is fixed (and the same) for all knots in that candidate surface.


This paper uses the principals of SALSA (for univariate splines) and the CReSS basis to present a spatially adaptive local smoothing algorithm for bivariate splines (SALSA2D). To demonstrate this approach we use carcass location data from Etosha National Park (ENP), Namibia to assess the spatial distribution of elephant deaths.  


The African Savannah Elephant (\textit{Loxodonta africana}) occurs across 37 African countries with Southern Africa holding the largest number of elephants on the continent \cite{thouless2016}. It is the largest living terrestrial mammal, social, intelligent, an ecosystem engineer, a species of great conservation concern, and has been studied extensively. However, continental African elephant populations are declining rapidly - so much so that in the 2021-1 IUCN Red List of Threatened species these elephants have been reclassified as Endangered \cite{iucnlist}. Reasons for this decline include poaching, habitat fragmentation and loss, unsustainable bushmeat harvesting, conflict with humans, and scarcity of food and water linked to frequent severe droughts or civil war \cite{chase2016, ripple2015}.

Elephant mortality is an important component of understanding the population dynamics, the overall increase or decline in populations and for disease monitoring.  Poaching has been the major focus of elephant mortality studies \cite{douglas1987, wittemyer2014, beale2018} with other causes such as human-elephant conflicts, accidents and natural processes (e.g. disease) less studied. With no natural predators, natural elephant mortality is often as a consequence of food scarcity and water stress during drought \citep{mukeka2022} or diseases such as Anthrax.  Recently, \citet{mukeka2022} highlighted that despite the considerable inter-annual and spatial variation in elephant mortality in Kenya, the impact of this variation on population dynamics has not yet been widely assessed.  



The Monitoring the Illegal Killing of Elephants (MIKE; \url   {https://cites.org/eng/prog/mike/index.php/portal}) programme is an international collaboration that collects and monitors trends related to the illegal killing of elephants from across Africa and Asia \cite{mike2018}. The MIKE project also seeks to monitor the effectiveness of field conservation efforts and is part of the Convention on International Trade in Endangered Species of Wild Fauna and Flora (CITES) initiative. MIKE operates in over 80 sites, across 43 elephant range states across Africa and Asia and rigorous protocols have been developed as part of this initiative to collect, analyse and build the capacity to better enforce the law and to reduce illegal elephant killings. The overall goal of MIKE is to provide information needed for elephant range states to make appropriate management and enforcement decisions, and to build institutional capacity within the range states for the long-term management of their elephant populations. In particular, they report PIKE (Proportion of Illegally Killed Elephants) for every range state. 



In contrast to much of Africa, there has been little to no poaching of elephants in Etosha National Park (ENP) and the population of elephants in Namibia is actually increasing \citep{Craig2020}. Despite the limited poaching activity in Namibia, the MIKE project has been active in ENP in Namibia for over a decade, and substantial resources are used to collect relevant abundance and mortality data by dedicated aerial surveys under strict survey protocols. As part of routine park activities, opportunistic data on carcass information is also recorded. Together these form the African elephant mortality database for ENP. To date these data have only been reported to the MIKE project and not analysed spatially.

Statistical modelling of these data is necessary since the park is very large ($\sim$23,000 km\(^2\)) and regardless of the survey regime, the observed counts will undoubtedly comprise a subset of a larger number of deaths. Reliable modelling results which accurately estimate the magnitude and location of elephant mortality in ENP are also not guaranteed and require the careful consideration of at least the following two points, 1) most wildlife, including elephant, rarely traverse a large salt pan: there is little vegetation to be found in the pan and the sometimes boggy terrain prevents travel for large animals such as elephant; 2) the spatial patterns of mortality are likely to be localised and patchy: the abundance of elephant in the park is far from homogeneous and the reasons for death (natural or otherwise) are also likely to vary across the park. Failing to account for the possibly unusual spatial patterns in these data and/or assuming points across the pan are as closely linked as equidistant points without a physical barrier, can unwittingly lead to false conclusions about the magnitude and location of elephant deaths in the park.

The Complex Region Spatial Smoother (CReSS) is a regression spline based statistical modelling method equipped to address both aspects of these data  \cite{scott2014}. Euclidean or geodesic (`around the salt pan') distances can be used to underpin the smoothed surface and the method is spatially adaptive enabling the targeting of surface flexibility to accommodate any particularly patchy trends and/or local surface features. While appropriate, the currently published CReSS method \cite{scott2014} undertakes the, crucially important, model selection process using a model-averaging of predictions approach which can be computationally intensive. We have also found after extensive use that this can mask unusually shaped spatial patterns when these are observed. In this paper, we propose using CReSS with an automated model selection approach, as an alternative to model-averaging, which enables atypical spatial patterns to be deduced from the data - patterns which have implications for park management in this case, and produces one model which is easier to handle.

In much of the grey literature the methods described here have been applied in a Poisson or Binomial generalised additive model framework (GAM).  Here we have chosen to showcase the versatility of the SALSA algorithms and apply them to a presence only data set, where the primary interest are the spatial locations of presence points (carcass locations).

This paper begins by describing the original CReSS method in more detail and then introduces the SALSA2D algorithm.  The first analysis presented focuses solely on spatial variation to compare the SALSA2D method to the original model averaging one. Lastly, using only the SALSA2D method for the spatial variation, environmental covariates are also added to the model to assess how carcass intensity varies with location, annual rainfall, distance to water and distance to roads.  

\section*{Methodology}

\subsection*{The Complex Region Spatial Smoother(CReSS)}\label{the-complex-region-spatial-smoother-cress}

The CReSS approach fits pure spatial regression models to a set of coordinates $\mathbf{x}$ of the form:

\begin{equation}\label{eqmodframework}
    g(\mathbf{y}) = \eta = \beta_0 + s(\mathbf{x})
\end{equation}

where $g$ is the link function and $\eta$ the linear predictor. $\mathbf{s}$ is a two dimensional surface approximated by a linear combination of exponential basis functions \(bE\). The formula for this basis function at observation $i$ and knot location $k$ is:

\begin{equation}\label{eqexp}
bE_{ki} = \exp^{(-h_{ki}/r_k^{2})}
\end{equation}

where \(r_k\) dictates the extent of the decay of this exponential function with distance between points, and thus the extent of its local nature. Notably \(h_{ki}\) indicates a geodesic or Euclidean distance (for some observation \(i\) and the \(k\)-th knot location). Parameter \(r_{k}\) takes values such that if \(r_k\) is small the model will have a set of relatively local basis functions and if \(r_k\) is large the model will have a set of relatively global basis functions. The exact values of \(r_k\) are dependent upon the range and units of the spatial covariates.

After the choice of distance metric, the CReSS with model averaging procedure fits multiple models with each model evaluated at one of a variety of parameter values for the number of knots (\(K\)) and the effective range parameter (\(r\)). According to \citet{scott2014} model selection is achieved using AIC\(_c\) \cite{sugiura1978} weights and averaging those models with \(\Delta\)AIC\(_c\) \(<\) 10 to produce weighted predictions.

We have also extended the suite of CReSS basis functions that can be used for the two-dimensional smoothing. This is useful since SALSA2D is agnostic about the basis function used but relies instead on an objective fit criteria for execution.

As part of recent work, we have expanded the CReSS approach to include a Gaussian radial basis to the choice of basis functions available for selection. The two bases have different shapes, with the exponential being more peaked at the centre.  These choices allow for more nuanced model fitting, akin to link function or distance metric choice. The Gaussian radial basis, \(bG\), is specified as:

\begin{equation}\label{eqgau}
bG_{ki} = \exp^{(-(h_{ki}r_k)^2)}
\end{equation}

where \(r_k\) and \(h_{ki}\) are as defined for the exponential (Equation \ref{eqexp}) except that for the Gaussian basis, a small value for \(r_k\) returns a relatively global basis and a large \(r_k\) value returns a relatively local basis.

The new basis and SALSA2D algorithm are all implemented inside the \texttt{MRSea} \texttt{R} package \cite{scott2017, Rcran2019} for easy use by practitioners.

\subsection*{Spatially Adaptive Local Smoothing Algorithm for at least two dimensions (SALSA2D)}\label{spatially-adaptive-local-smoothing-algorithm-for-at-least-two-dimensions-salsa2d}

SALSA2D uses the same model framework as for model averaging (see Equation \ref{eqmodframework}) but where the knot locations, $k$, are chosen using an iterative three step procedure. The algorithm works in (at least) two dimensions and begins with space-filled knots to facilitate spatial coverage and then adaptively moves, adds and drops knots into, or from, locations in line with poor model fit (evidenced by large residuals) and an objective fit criteria. At each stage, the global/local extent of each basis function via the \(r_k\) value employed can also be revised as part of the search for a more appropriate surface. So, unlike the model averaging approach, SALSA2D returns one model with specifically selected $k$ and $r_k$ enabling standard methods for assessment of fit and uncertainty estimation. 

The algorithm that drives SALSA2D has an iterative 3-step structure. After an initialisation step, there are three repeated steps: the first is a simplification step to reduce the number of estimated parameters which is achieved by allowing for the removal of columns from the design matrix (reduction in knot number). The second and third steps (exchange and improve) are designed to efficiently search the model space (all possible number and locations of knots). The exchange step allows for the possibility of moving away from a local optimum or addition of columns to the design matrix (a new knot) and the improvement step attempts to make local improvements in knot location. The outcome of each of these steps is determined by an objective fit criterion and repeated until no improvements are made (or an iteration limit is reached). The structure of the algorithm is given in the pseudo code in Figure 1 and the next sections describe the steps in detail.

\begin{figure}[!htb]
 \label{fig:salsaAlg}
\vspace{1cm}

\fbox{\parbox[c]{\linewidth}{
SALSA2D:\\
Given an $n$-dimensional set $K_l$ of possible knot locations over the region of interest.\\
\hspace*{2em}\textit{Initialise}\\
\hspace*{2em}\hspace*{2em}Initialise knots, $K_s$ within the points of $K_l$\\
\hspace*{2em}\hspace*{2em}Check for convergence\\
\hspace*{2em}\textit{Repeat}\\
    \hspace*{2em}\hspace*{2em}\textit{Repeat} Simplification step \textit{while} ($K>K_{\textrm{min}}$ and fit measure improves)\\
    \hspace*{2em}\hspace*{2em}\textit{Repeat} Exchange step \textit{while} ($K<K_{\textrm{max}}$ and fit measure improves)\\
    \hspace*{2em}\hspace*{2em}\textit{Repeat} Improvement step \textit{while} (fit measure improves)\\
    \hspace*{2em}\textit{While} (an improvement in fit measure is made by one of the above steps)\\
   
}}

\caption{Pseudo-code outlining the structure of SALSA2D \cite[adapted from Figure 1,][]{Walk2010}, where $K$ is the number of knots used for fitting.}
\end{figure}

\subsubsection*{Initialisation}\label{initialisation}

Each observed location, \(i\), is considered a possible location for a knot position. To avoid estimation issues, only unique knot locations are considered giving \(K_l\) legal knot locations. The user specifies a starting number of knots, \(K_s\), where \(K_s<K_l\), and these are selected from \(K_l\) using a space-filling algorithm \cite{Johnson1990}. This method provides good coverage across the spatial region as a starting position for SALSA2D. Additionally, the minimum number of knots, \(K_{\textrm{min}}\) (\(2\leq K_{\textrm{min}}<K_s\)) and maximum number \(K_{\textrm{max}}\) (\(K_s< K_{\textrm{max}}\leq K_l\)) are specified.

To evaluate the basis function, the \(r_k\)-value for each basis must also be chosen. The SALSA2D algorithm selects from \(R\) possible options for \(r_k\) which range from a very local basis to a globally acting basis. The middle option which is neither very local or very global, is chosen to initialise the first model.

To ensure that the initial model fit has converged, there is a drop step component that is activated if the variance of the initialised first model exceeds that of the simpler input model (the variance should not increase with additional parameters/flexibility in the model). If this occurs, knot locations with the largest contributions to the variance are removed one by one until the overall variance of the more complex model is lower than the input model.

\subsubsection*{The simplify step}\label{the-simplify-step}

Using the fit criteria specified, the simplify step compares the current model with all models obtained by removing an existing knot (as long as this is at least \(K_{\textrm{min}}\)). At each iteration, the model with the best fitness measure is retained and the process repeated until there is no further improvement in the fitness measure. This step can be carried out by fixing \(r_k\) or by choosing \(r_k\) for each basis as each knot is dropped for comparison.

\subsubsection*{The exchange step}\label{the-exchange-step}

The exchange step increases the extent of the search of model space by enabling a move away from a local minima (of the fit criterion). It uses the maximum Pearson residual from the current fitted model to identify a possible candidate location for a new knot (although in theory other types of residuals could be chosen and we use an alternative metric for the point process models in the next sections). The algorithm then compares the objective fit criteria for these models that result when each of the existing knots in the current model is moved to this new location, and also the fit criteria from the model that results when an additional knot at this location is added to the current model (if this does not exceed \(K_{\textrm{max}}\)). The model with the best fitness measure is retained in this step if it has a better fitness measure than the current model. Evaluation of each of these models can be very quick to return but this process is naturally more computationally expensive, if \(r_k\) is also chosen for each basis function for each candidate model. In practice, the algorithm uses the knot locations of the five largest residuals as candidates for an exchange or move.

\subsubsection*{The improve step}\label{the-improve-step}

The improve steps allows a more nuanced search of the local minima by allowing small adjustments to the location of each knot. Using the fit criteria specified, the improve step compares the current model with all models obtained by moving an existing knot to one of its five nearest neighbours (determined by the distance metric employed: geodesic or Euclidean). At each iteration, the model with the best fitness measure is retained. As with the exchange step, alternative choices for the \(r_k\) parameter may be considered when fitting each new model and this process is likely to be swift at this stage.

\subsubsection*{Determining $r_k$}
\label{determining-r_k}

This routine considers incrementing or decrementing \(r_k\) values in the sequence of \(R\) possible values, where the sequence is selected using the method from \citealp{scott2014}. It can be evaluated either once at the end of the exchange, improve and simplify steps or as part of every decision taken during these steps. The process is done by considering each of the radial basis columns in turn, and incrementing or decrementing the \(r_k\) values in the index until there is no improvement in the fitness measure. At each step the \(r_k\)-values for the other basis columns are maintained at the current solution. The best of these models is selected as the new current model, and the process iterates until no improvement is made. This process can have a large computational overhead and may significantly prolong the procedure but constitutes a broader search of the model space. 

This algorithm is implemented in the \texttt{MRSea} package which can be found at \url{http://lindesaysh.github.io/MRSea/} \citep{scott2017}.

\section*{Methods comparison}\label{methods-comparison}

This section compares the performance of the CReSS with model-averaging approach to CReSS with SALSA2D for model selection. The methods are compared numerically using log-likelihood, while the practical consequences of using each are assessed visually and contextualised using surface features in Etosha National Park, Namibia.

\subsection*{Data specification}\label{data-specification}

To appreciate the numerical and practical benefits of this methodological development, the MIKE data was used for the method comparison and for the subsequent analysis in full \cite{mike2018}. These data consists of 320 carcass locations observed between February 2000 and March 2017 in Etosha National Park (ENP). The observed fatalities are recorded as being due to: anthrax, natural (age-related) causes, poaching and unknown. While a substantial proportion of the carcasses are recorded as being for `unknown' reasons (54\%) the largest known cause of death is from Anthrax (27.8\%; 12.5\% confirmed cases and 15\% suspected). Less than 1\% of the carcasses were confirmed as poached. All reported carcasses were used in the analysis regardless of type. Disregarding 2017 as it was only a partial year, 2006, 2013 and 2014 had the fewest recorded carcasses (7-8), whilst 2002, 2003, 2005 and 2011 had the highest recorded (27-28). As there were a relatively small number of observations per year, no guarantee the deaths occurred in the year of detection and no obvious changes in the spatial pattern of observations, the data were pooled across years. 
  
\begin{figure}[!htb]
\centering
\includegraphics[width=0.9\linewidth]{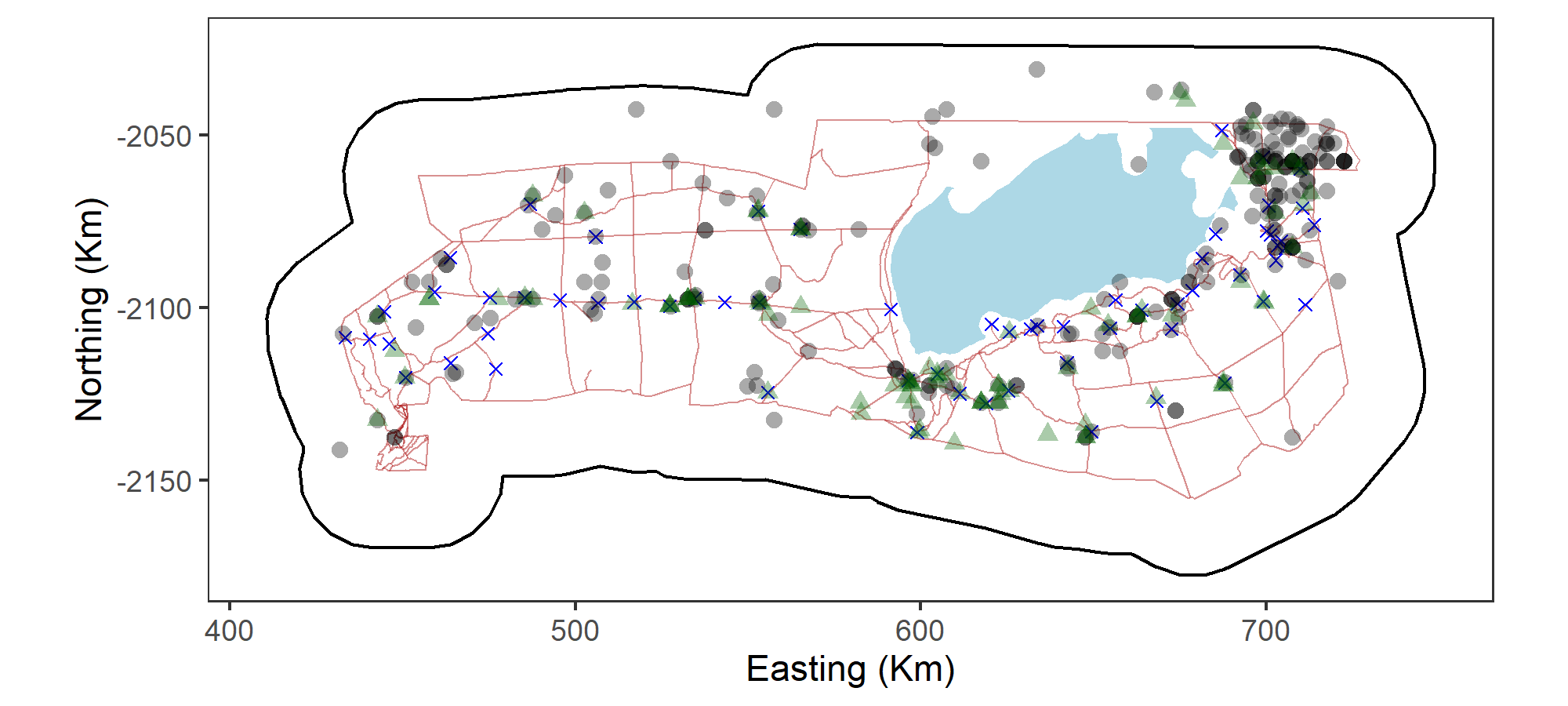}
\caption{Figure showing the study area (Etosha National Park) with the carcass locations. The  green triangles show confirmed or suspected anthrax cases and grey/black all other types. As there are duplicate locations, the darker the shape, the more presence locations. The study area (park boundary plus 20km buffer) is outlined in black. The blue polygon is the Etosha salt pan, the red lines are park roads and the blue crosses are waterholes. The outermost red line is also the park fence.}
\label{fig:raw}
\end{figure}

The longitude and latitude coordinates were converted to Universal Transverse Mercator (UTM) zone 33S and the study region was extended beyond the ENP boundary by 20 km to allow for the inclusion of carcasses just outside the park. Additionally, the large salt pan was reduced in size by 2 km to allow inclusion of carcasses found near the edge of the pan.  The data show that carcasses generally seem to occur near roads (or, at least, are more commonly observed near roads) and waterholes (Figure \ref{fig:raw}). It is possible that these patterns are due to opportunistic reporting of carcasses as a result of park vehicles moving along the roads, however the data were from both opportunistic and dedicated surveys, which are carried out without reference to roads. Furthermore, collared elephant in ENP have been shown to utilise roads/tracks and fire breaks extensively and are known to frequent waterholes \citep{Tsalyuk2019, chamaille2007}. Deaths as a result of anthrax appear to be particularly well correlated with waterholes (Figure \ref{fig:raw}).

In this data set, the link back to the original survey effort (where surveys were undertaken) is not available so we are left with only the carcass locations and no absence locations. \citet{warton2010} showed the link between logistic regression and an inhomogeneous Poisson point process model (PPM) and here we use both this link and the downweighted Poisson regression method \cite{renner2013} to fit a Poisson PPM using a pure regression GAM framework. In this case, the intensity is the number of presence records (carcass sightings) per unit area and is modelled as a function of covariates measured throughout the study region. It is a relative measure and gives the expected abundance of carcass sightings for a given area. 

Pseudo-absences in the regression setting play the same role as quadrature points in point process modelling and we used the point process framework to choose the number and location of these points.  The pseudo-absence points were selected as a regular grid and the number based on convergence of the likelihood \cite{renner2013}.

Lastly, to determine areas of poor fit, the exchange step requires the calculation of residuals.  This was achieved by creating a neighbourhood around each knot location ($k$) and comparing the observed number of points with the sum of the estimated intensities in the same area. For more details, see Section 1 of the Appendix S1.

\subsection*{Model specification}\label{model-specification}

To compare the performance of SALSA2D with model averaging as a model selection approach, models with a two dimensional smoother-based term for geographic locations were fitted to the MIKE data. The comparison involved either the published CReSS method which employs model averaging \cite{scott2014} or model selection using SALSA2D to determine knot number and location. Here we model the the locations of the carcasses jointly with the pseudo-absences by maximising the following weighted Poisson log-pseudolikelihood \cite{Berman1992}:

\begin{equation}\label{eq:loglik}
    l(\beta;\textbf{X}) = \sum_{i=1}^{N}w_i(y_i \textrm{log}(\lambda(\textbf{X}_i)) - \lambda (\textbf{X}_i))
\end{equation}

where \(\lambda(\textbf{X}_i))\) is the intensity at location \(i\),  $\boldsymbol{X}_i$ represents the design matrix at location $i$, $N$ is the total number of points (presence and pseudo-absence), $\textbf{w} = \{w_1, \cdots, w_N\}$ are quadrature weights.

\begin{align*}
    y_i = \begin{cases}
      \frac{1}{w_i} & \text{if $i$ is a presence location}\\
      0 & \text{if $i$ is a pseudo-absence location}
    \end{cases} 
\end{align*}
 
 The log-pseudolikelihood in Equation \ref{eq:loglik} \cite{Berman1992} is a re-expression of the Poisson PPM log-likelihood \cite{cressie1993}, which means that models can be fitted using standard GLM software.  Here we model the expected number of carcasses per square kilometre and so the weights for the pseudo-absence points are specified as the area of the study region, 37,872 km$^2$ (ENP plus the 20 km buffer) divided by the number of pseudo-absences. The weights for presence points are set to some small value ($10^{-6}$).

Likelihood convergence was used to determine the the number of pseudo-absences which was estimated to be 9644 (a grid spacing of 2 km). For more details see Section 2 of Appendix S1. 


For this method comparison section, we model the intensity as a function of coordinates, $\mathbf{x}$, only.

\begin{equation}\label{eq:modcompeq}
\textrm{log}(\lambda(\textbf{X}_i)) = \eta_{i} = \beta_0 + s(\mathbf{x})=\boldsymbol{X}_i^T \boldsymbol{\beta}
\end{equation}

where \(\eta_i\) is the linear predictor, consisting of the intercept, \(\beta_0\), and a smooth function of coordinates, s($\mathbf{x}$). The smooth function is either the exponential or Gaussian basis function.

For both the model averaging and SALSA2D methods, the following specifications were used to return the columns of the design matrix \(\boldsymbol{X}\) in Equation \ref{eq:modcompeq}:

\begin{itemize}
\item
  Two basis options: Exponential (\(bE_{ki}\); Equation \ref{eqexp}) or
  Gaussian (\(bG_{ki}\); Equation \ref{eqgau})
\item
  Two distance measures (Euclidean or geodesic) to calculate $h$ in the basis equations; the geodesic distances are calculated using Floyds algorithm \cite{Floyd1962} and for more details see \cite{scott2014}. In this study, geodesic distances are ``around the salt pan'' distances. 
\item
  12 choices of fixed knot number (for the model-averaging approach) and 12 choices of starting knot numbers, \(K_s\) for the SALSA2D approach. In each case, the fixed/starting knot set was: {[}5, 10, 15, \dots, 55, 60{]}. A total of 285 legal knot positions ($K_l$) were considered. These consisted of all non-duplicated carcass locations (n=245) and 50 space-filled pseudo-absence locations ($\sim20$\% of all $K_l$).
\item
  10 choices of \(r_k\) (also specified as part of Equations \ref{eqexp}
  and \ref{eqgau})
\end{itemize}

Additionally, for SALSA2D, \(K_\textrm{min}\) and \(K_\textrm{max}\) were set to 2 and 100 respectively, for all model specifications.

\subsection*{Model comparison}\label{model-comparison}

In keeping with \citet{scott2017}, the model-averaging CReSS method was governed by AIC\(_c\) weights which were used to choose which models to average (\(\Delta AIC_c \leq10\)) and their relative contribution to the overall averaged model. In keeping with \citet{Walk2010}, the BIC was used to govern SALSA2D model selection regarding the choice of knot number and their locations across the range of combinations of basis type, distance metric, starting knot number and \(r_k\) choices \cite{schwarz1978}.  In all cases, the log-likelihood score (Equation \ref{eq:loglik}) was calculated for each model to enable comparison between model selection strategies. 







\section*{Results}\label{results}

\subsection*{Numerical comparison}\label{numerical-comparison}

The log-likelihood scores returned for the model averaging method were fairly close (maximum difference 14 points) regardless of the basis function and distance metric used in each model (Table \ref{tab:mrsearesult}, Method: `Model averaging'). The geodesic-exponential combination scored the best (largest log-likelihood) of the 4 combinations trialled. Interestingly, this combination chose 11 models with which to average over to obtain this solution, compared with some options that chose far fewer models to use as part of the average calculation. In general, geodesic distances were preferred to Euclidean regardless of basis. 

\vspace{1cm}

\begin{table}[!ht]
\centering
\caption{Table showing the results of the model averaging and SALSA2D methods of model selection for a given basis type and distance metric used. The `No. Models' indicates the number of models chosen to carry out the model averaging in each case, and the `No. Knots' indicates the number of knots chosen for each model using the SALSA2D selection method. The star indicates the model with the largest log-likelihood (LL) score, and thus the chosen model based on the LL in each case.}
\vspace{0.5cm}
\begin{tabular}{l |l|l|c|c|c}
Method & Basis & Distance Measure & No. Models & No. Knots & Log-Likelihood\\
\hline
MA & Exponential$^*$ & Geodesic & 11 & - & -1432.0\\
 &  Gaussian & Geodesic & 2 & -& -1441.5\\
 & Exponential & Euclidean & 1 & -& -1443.4\\
 &  Gaussian & Euclidean & 8 & - & -1446.3\\
 \hline
SALSA2D & Exponential & Geodesic & - & 32 & -1369.7\\
& Gaussian & Geodesic & - & 32 & -1408.3 \\
& Exponential$^*$ & Euclidean & - & 41 & -1301.6\\
& Gaussian & Euclidean & & 47 & -1541.6\\
\end{tabular}
\label{tab:mrsearesult}
\end{table}

 The log-likelihood scores for the SALSA2D based selection are shown for the model with the highest log-likelihood for each of the basis/distance metric combinations (Table \ref{tab:mrsearesult}, Method: SALSA2D). Across the four combinations, the scores were less homogeneous than for the model averaging results and the exponential-Euclidean SALSA2D model (using 41 knots) was the best of all trialled here. In contrast to the averaging approach, there was a preference for the exponential basis with the distance metric secondary. In reality, the user may prefer to select the best model using BIC (as was used for $k$/$r$ selection). In this case, the order of the four parameterisations was the same (exponential-Euclidean the best and Gaussian-Euclidean the worst) and the best model using BIC was the same as in Table \ref{tab:mrsearesult} when log-likelihood was used (see Section 3 of Appendix S1 for an expanded version of Table \ref{tab:mrsearesult}).

\begin{figure}[!ht]
\centering
\includegraphics[width=0.7\linewidth]{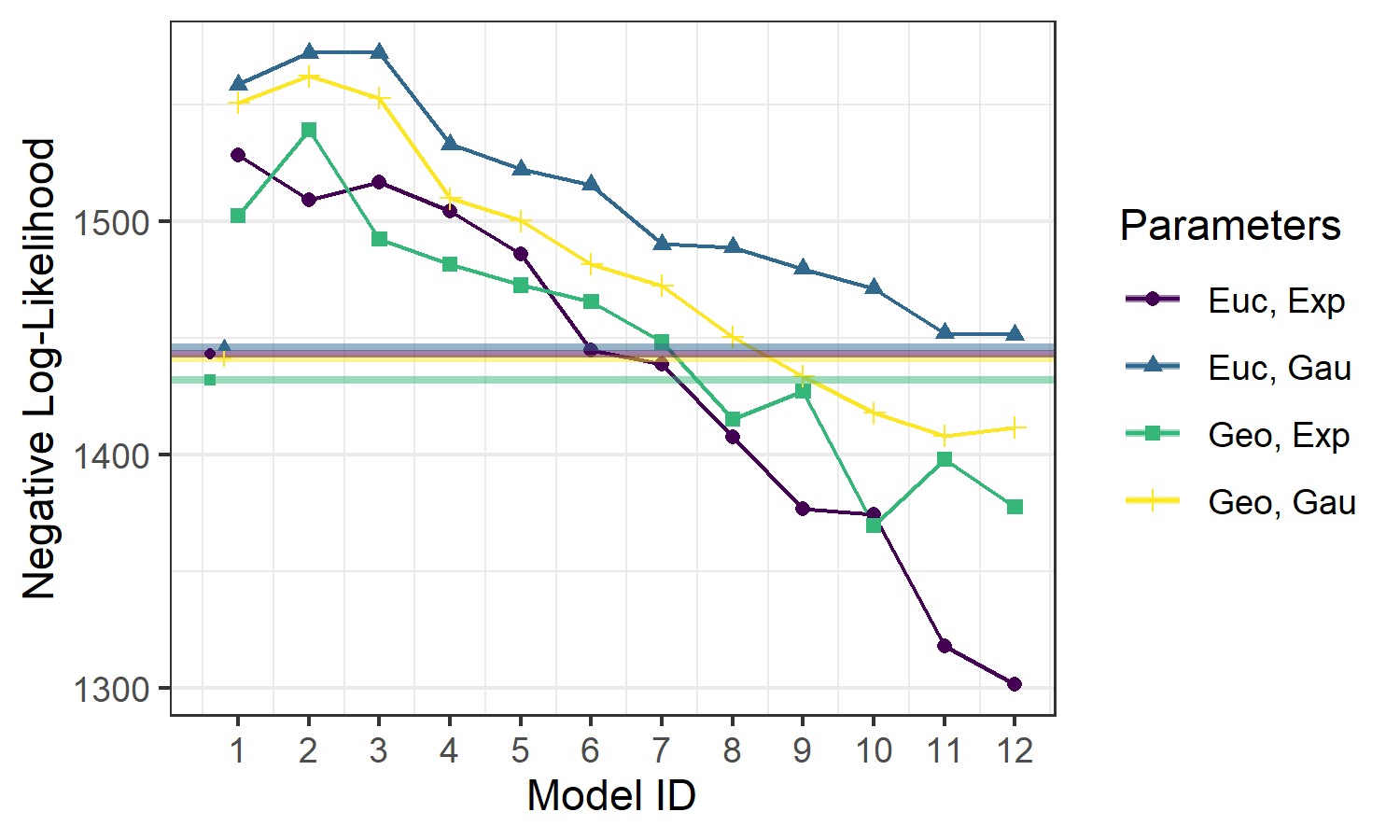}
\caption{Figure showing the model identification number (increasing start knots) and the negative log-likelihood score for each of the SALSA2D models resulting from a different start knot number, $K_s$. The horizontal lines are the scores for the equivalent model averaging result. (Euc - Euclidean, Geo - Geodesic, Exp - Exponential and Gau - Gaussian)}
\label{fig:simresults}
\end{figure}

Using the ``best'' SALSA2D models only, for all but one combination of basis type and distance metric used, all SALSA2D models produced better scores than the model averaging method -- sometimes reducing the log-likelihood score by as much as 10\%. However, if SALSA2D initialises with too few knots, the algorithm may get stuck in local minima. So long as a large enough number of starting knot locations was selected ($\sim\geq 40$), SALSA2D-based selection resulted in superior scores over the model-averaging alternative (Figure \ref{fig:simresults}). This demonstrates that the SALSA2D model selection method can return improved results and at worst, SALSA2D results were almost indistinguishable from the best model averaging-based result.

\subsection*{Visual comparison}\label{visual-comparison}

\begin{figure}[!ht]
\centering
\includegraphics[width=0.7\linewidth]{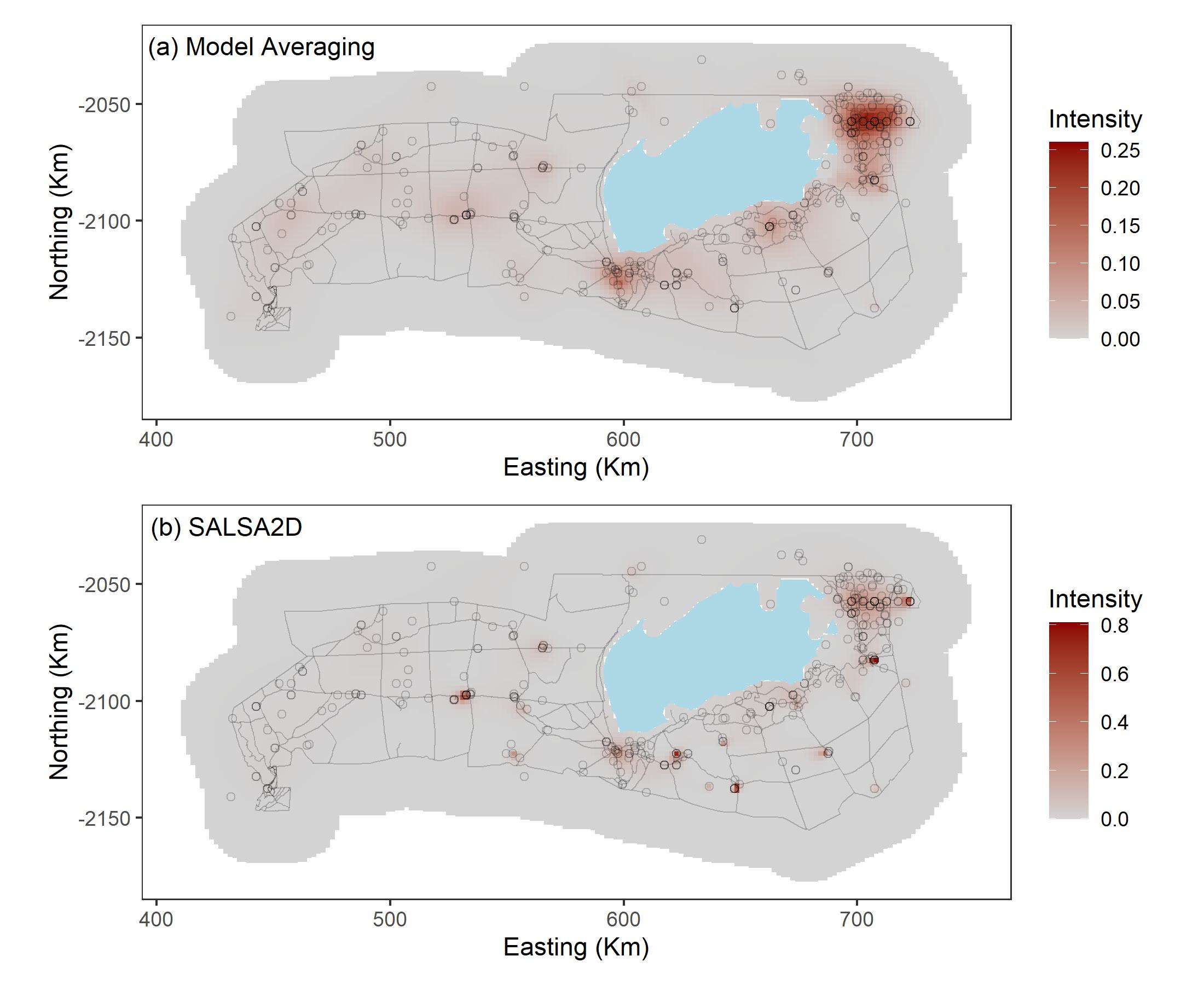}
\caption{Figure showing the intensity of carcass locations throughout Etosha using the model averaging (top) and SALSA2D (bottom). Note: to ensure detail can be seen, the two images have differing intensity scales. The carcass locations are shown as black circles. The blue polygon is the Etosha salt pan and the lines are the roads within the park.}
\label{fig:simmaps}
\end{figure}

Results for the model-averaging based model (Figure \ref{fig:simmaps}a) signalled that the intensity of carcasses was highest in the north-east of the park (where most the observed deaths occurred) and along the southern edge of the large salt pan, which is consistent with the observed data. The carcass intensity is very low near the south-west and south-eastern borders of ENP.

Whilst the model averaging results show a smooth intensity surface, the SALSA2D method produces a more clustered intensity surface (Figure \ref{fig:simmaps}b). The surface shows more local effects, particularly the centre west and below the salt pan and the highest intensity at these spots was nearly three times that of the model-averaging result. These effects, match well with the carcass location data and frequently occur at the confluence of several roads and some waterholes. 


\begin{figure}[!ht]
\centering
\includegraphics[width=0.7\linewidth]{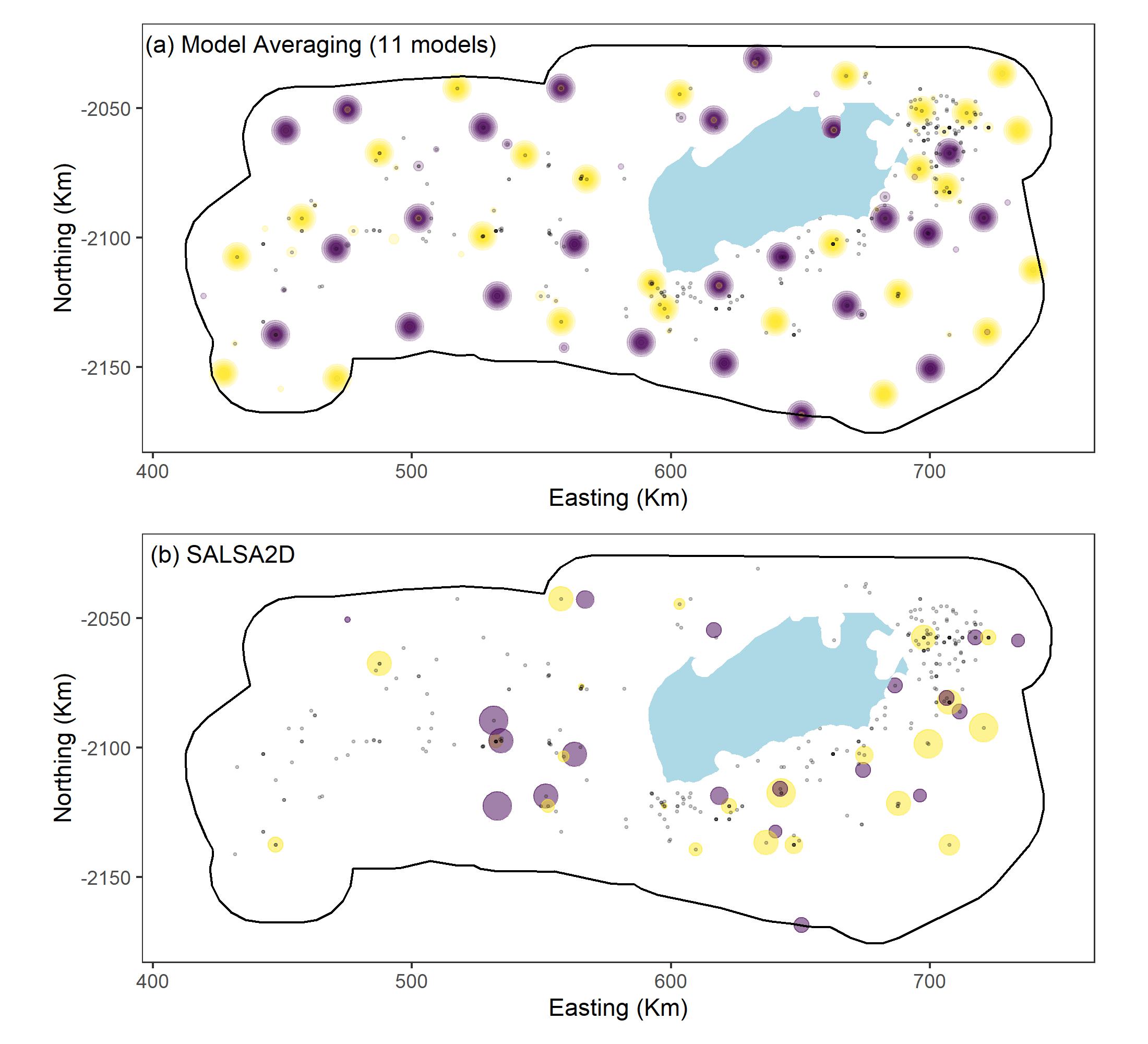}
\caption{Figure showing the knot locations and $r$ (effective range of basis function) from the best model averaging (top) and SALSA2D (bottom) models. Yellow is for a positive model coefficient and purple a negative one. The size of the coloured circles is a visual representation of the size of the $r$ parameter. Note that in (a) the concentric rings are from models had the same knot locations with different $r$. In (b) the colours overlap but each $k$ is in a different location.  The carcass locations are shown as grey/black circles. The blue polygon is the Etosha salt pan.}
\label{fig:simknots}
\end{figure}

Figure \ref{fig:simknots} shows the selected knot locations and equivalent $r$ parameter from the 11 averaged models (Figure \ref{fig:simknots}a) and the one best SALSA2D model (Figure \ref{fig:simknots}b). The averaged knot locations are more difficult to represent but it can be seen that there are multiple $r$ values (ranging from global to very local) across the same locations and occasionally a location where the sign of the coefficient changes between models. The SALSA2D result is more nuanced with very few knot locations selected to the west of the park. For the 41 selected locations, a variety of $r$'s were chosen. It is interesting that the SALSA2D approach found the Euclidean distance metric to be best and it is possible that the more local knots chosen under this method negate the need for the geodesic distances by limiting the possible leakage across the pan.  

\clearpage

\section*{Application: Analysis of Elephant mortality in Etosha National Park}\label{analysis-of-elephant-mortality-in-etosha-national-park}
Model
\subsection*{Data available}\label{data-available}

The intensity of elephant carcasses, based on the observed carcass locations and pseudo-absences, was modelled using four candidate covariate terms: distance from the nearest road, distance from the nearest water point, mean annual rainfall and a spatial term based on spatial coordinates (in km, UTM zone 33S).

The distance from nearest road and nearest waterhole metrics were calculated using shape files supplied by the Ministry of Environment, Forestry and Tourism (Namibia). These metrics were considered as candidates in the model to reflect possibly differential mortality rates near roads and waterholes, regardless of their spatial location in the park. 

The mean annual rainfall was based on rainfall data collected from 168 rain gauges distributed across Etosha National Park which are visited annually, when possible. Annual rainfall was not available for every gauge for every year, due to logistical difficulties reaching remote areas in some years, and so this metric was averaged across years for each gauge before interpolation to indicate areas in ENP with persistently high or low rainfall. The interpolation was achieved using a high dimensional penalised spline (\(df=150\)) to allow for interpolation to the carcass data locations and to the pseudo-absence grid. Details on the rainfall interpolation can be found in Section 4 of Appendix S1.

The proximity to waterholes was included as a candidate since elephant frequent water holes throughout the year, particularly in the dry season; roughly May to October \cite{Tsalyuk2019} and have been shown to have increased habitat use with proximity to water. \cite{Harris2008}.

While natural deaths might occur in line with their distributional patterns it is thought Anthrax-related deaths may be related to the use of water holes \cite{zidon2017}. The relationship with waterholes was found to be very stepped and so this variable was converted to a 2 level factor; $<3$km and $\geq 3$km (cutoffs of 1-5km were trialled and assessed using BIC).

The reasons for including proximity to roads as a candidate might seem less obvious, but the attraction or repulsion to roads by elephants might also be evident in their mortality patterns, and the model comparison work demonstrated that some roads are important (Figure \ref{fig:simmaps}b). This could be due to elephant preference to be found near roads, which is possible owing to their extensive use of roads/tracks for travel \cite{Tsalyuk2019}, but can only be confirmed by a dedicated analysis of survey data or that the detection of carcasses is higher near roads (e.g.~easier to observe).

The spatial term was considered for inclusion in this model to represent the spatial patterns in mortality that are not adequately explained by proximity to roads, water holes or annual mean rainfall. The role of this term in this model is crucial in this case - correctly identifying systematic spatial patterns in mortality might provide insights about other park features not currently considered to be related to mortality and overlooking these features prevents the mitigation of future elephant mortalities, particularly those related to poaching.

\subsection*{Model specification}\label{model-specification-1}

We are interested in modelling the intensity of elephant carcass locations as a function of distance to water, roads, mean annual rainfall and as a spatially adaptive smooth function of spatial coordinates. The model specification was:

\begin{equation*}
\begin{split}
\textrm{log}(\lambda(\textbf{X}_i)) & = \eta_{i}\\
& =\beta_0 + \textrm{distWater}_i + s_1(\textrm{rainfall}_i)\\
& \hspace{0.5cm} + s_2(\textrm{distRoads}_i) + s_3(\mathbf{x})\\
&  = \boldsymbol{X}_i^T \boldsymbol{\beta}
\end{split}
\end{equation*}

In this case, \(\lambda(\textbf{X}_i)\) is the intensity at location $i$ and $\textbf{X}_i$ represents the coordinates and environmental covariates. $s_1$ and $s_2$ represent one-dimensional basis functions, while $s_3(\mathbf{x})$ represents a two-dimensional exponential basis function for the spatial coordinates. $\boldsymbol{\beta}$ is a vector of model parameters associated with all columns of the design matrix, $\mathbf{X}$. The columns of $\mathbf{X}$ comprise the intercept (1), water $\geq3$km (0,1), $B$-spline bases for rainfall and roads and the exponential radial basis for the spatial term. 

Specifically, quadratic $B$-splines with SALSA based knot selection \cite{Walk2010} were used to implement the one dimensional smooth terms for rainfall and roads. The two-dimensional spline basis function was determined using Equation \ref{eqexp} (exponential basis) and based on Euclidean distances. Knot number, their locations and \(r_k\) values were chosen using the SALSA2D algorithm. The starting parameters were based on the best result from the simulation study; \(k_s = 41\), \(k_{\textrm{min}}=2\) and \(k_{\textrm{max}}=100\). The BIC was used to govern model selection in all cases.

\subsection*{Results}\label{results-1}

\begin{figure}[!ht]
\centering
\includegraphics[width=0.7\linewidth]{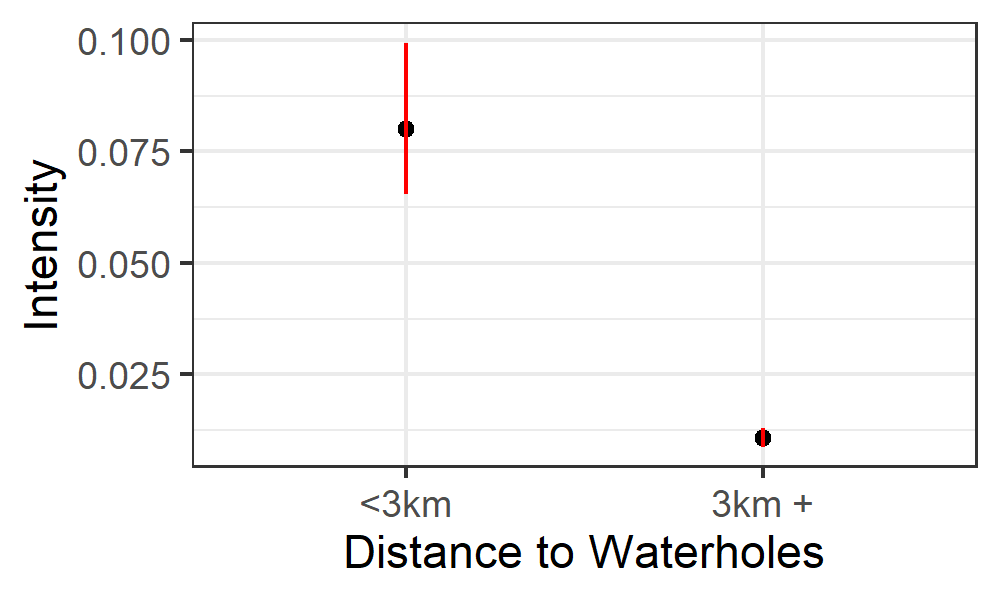}
\caption{Figure showing the estimated relationship of distance to the nearest waterhole to carcass intensity (when distance to roads = 1.5km and mean annual rainfall = 420mm).  The red line area is a 95\% confidence interval about the estimated relationship.}
\label{fig:mikeresultpartialwater}
\end{figure}

\begin{figure}[!ht]
\centering
\includegraphics[width=0.7\linewidth]{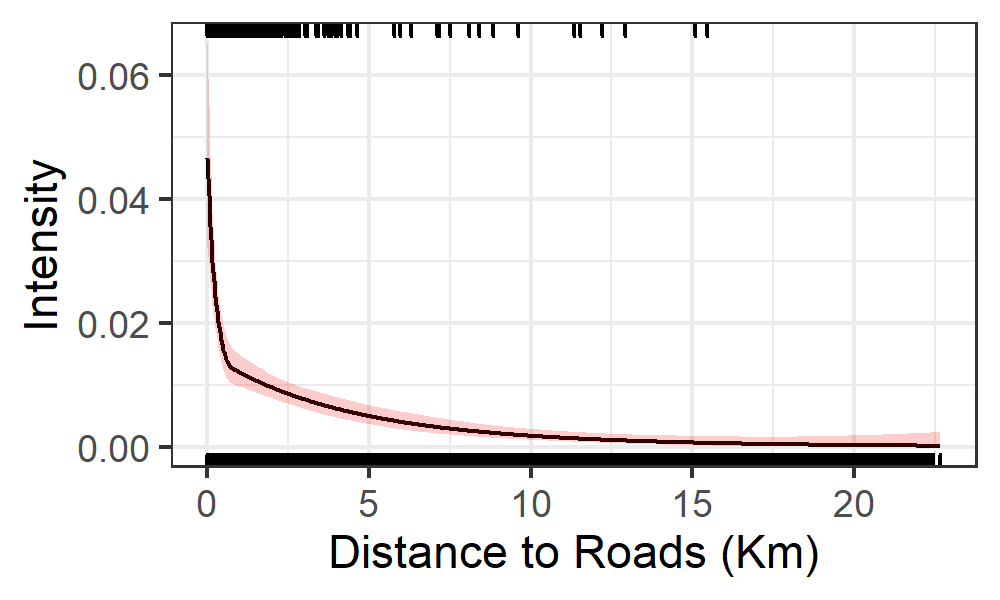}
\caption{Figure showing the estimated relationship of distance to roads to carcass intensity (when mean annual rainfall = 420mm and distance to waterhole is $\geq3$km).  The red shaded area is a 95\% confidence interval about the estimated relationship. The tick marks top and bottom show the values of the covariate in the original data which were presence locations (1's) and absences (0's).}
\label{fig:mikeresultpartialroad}
\end{figure}

\begin{figure}[!ht]
\centering
\includegraphics[width=0.7\linewidth]{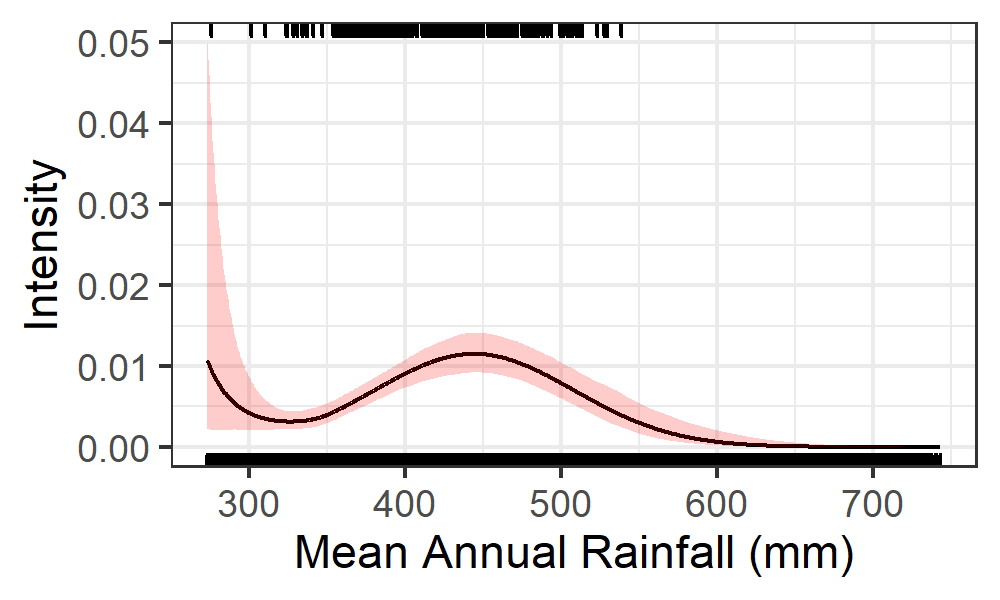}
\caption{Figure showing the estimated relationship of mean annual rainfall to carcass intensity (when distance to road = 1.5km and distance to waterhole is $\geq3$km).  The red shaded area is a 95\% confidence interval about the estimated relationship. The tick marks top and bottom show the values of the covariate in the original data which were presences (1's) and absences (0's).}
\label{fig:mikeresultpartialrain}
\end{figure}

The results show that carcass intensity is highest near to water holes and roads (Figures \ref{fig:mikeresultpartialwater} \& \ref{fig:mikeresultpartialroad}) and locations where the annual rainfall is approximately 450mm (Figure \ref{fig:mikeresultpartialrain}). Specifically, intensity decreases steeply with the distance from road until approximately 1km when the relationship subsides. 

The addition of distance from roads and mean annual rainfall to the spatial term, improved model results when compared with model results based on a SALSA2D-based spatial term alone (Models 2 and 3 Table \ref{tab:1Dresult}); the BIC scores substantially improved from 2980 to 2848.

\begin{figure}[!ht]
\centering
\includegraphics[width=\linewidth]{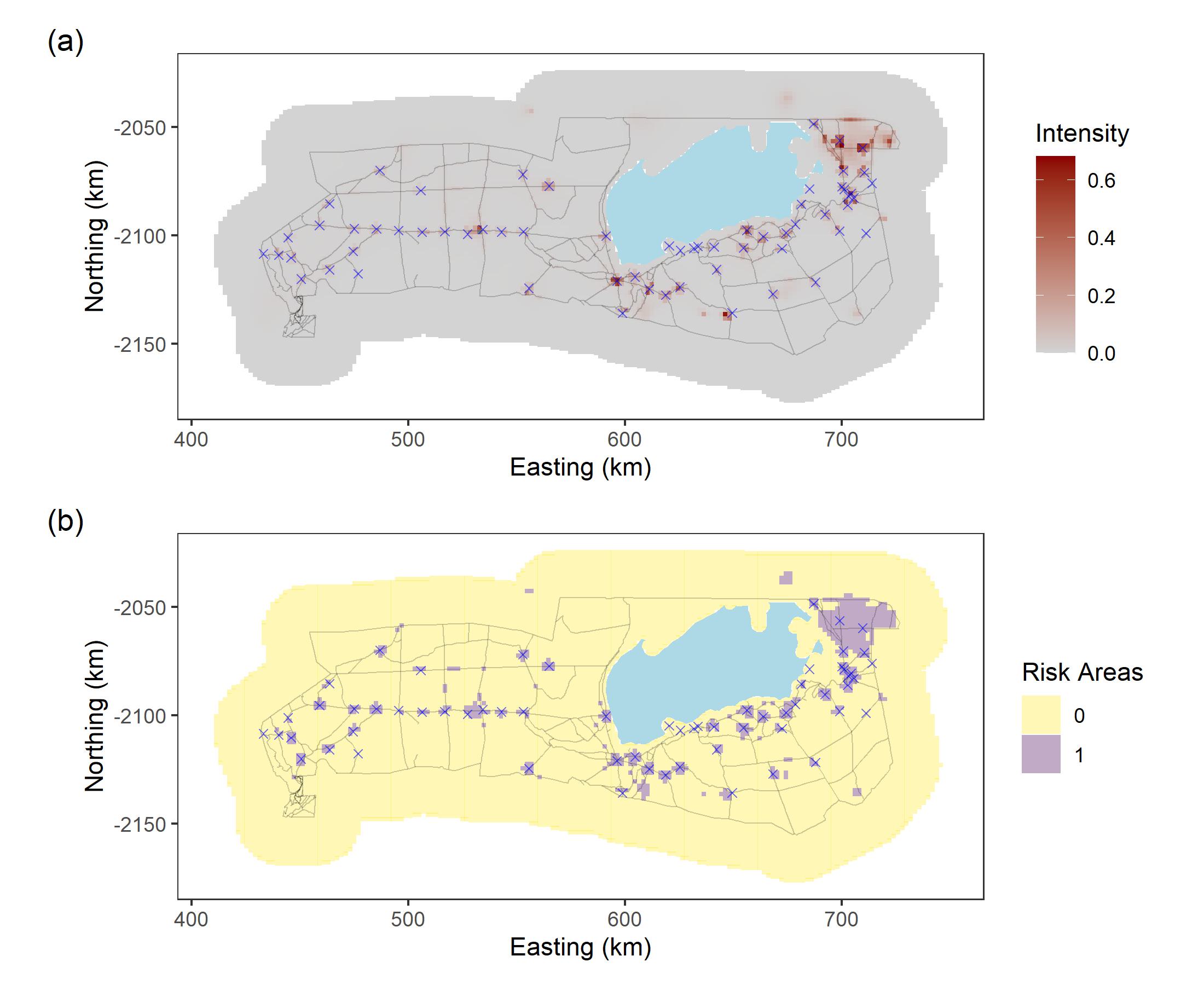}
\caption{Figure showing the estimated carcass intensity throughout the study area using SALSA and SALSA2D-based model selection and both one and two dimensional spline based terms (a). Figure showing the top 5\% intensity areas. The carcass locations are shown as black circles, the blue polygon is the Etosha salt pan, the blue crosses are waterholes and the black lines are roads.}
\label{fig:mikeoutputs}
\end{figure}

\begin{figure}[!ht]
\centering
\includegraphics[width=\linewidth]{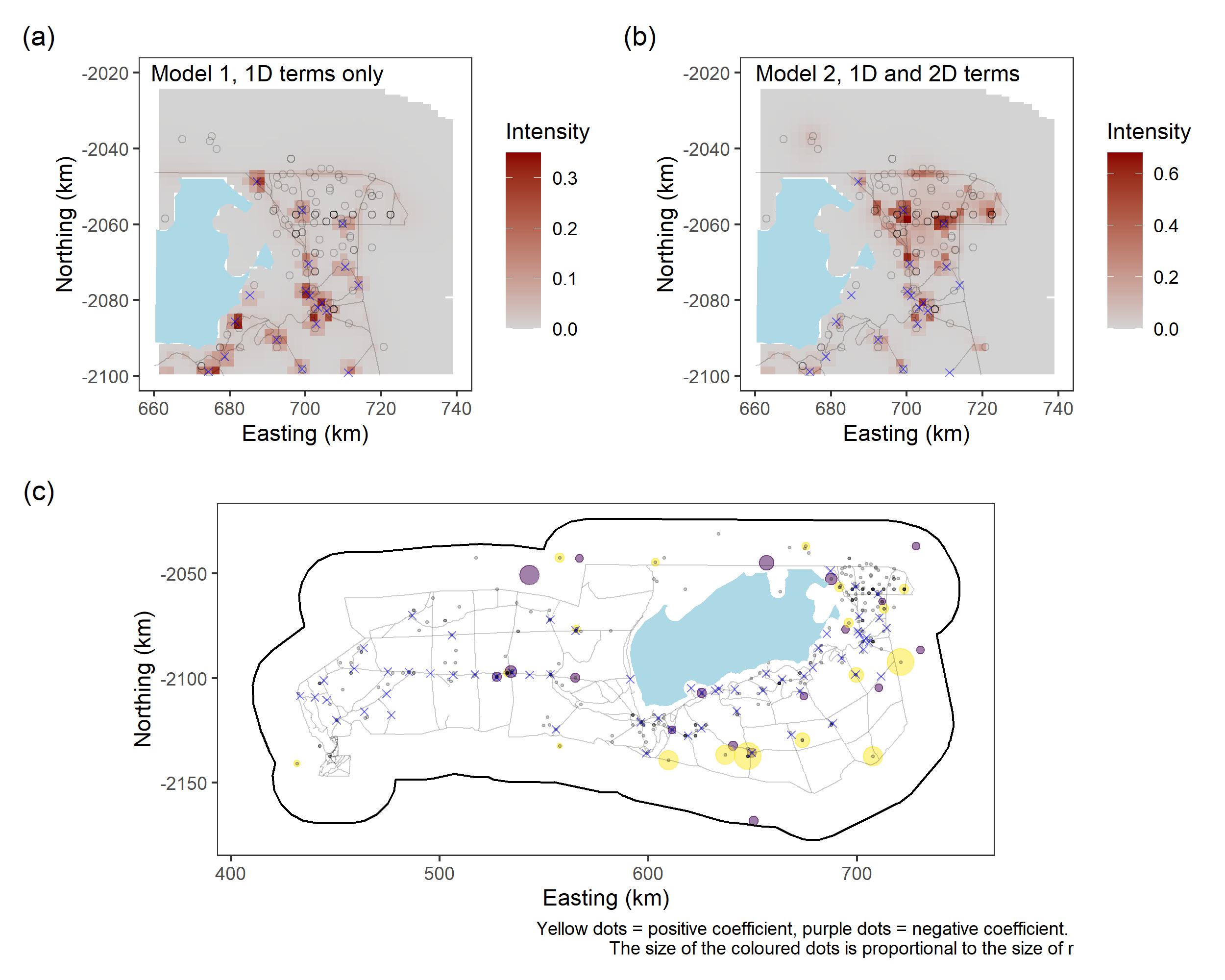}
\caption{Figure showing the estimated carcass intensity for the north-east part of the study area for (a) Model 1, (b) Model 2 and (c) the location of $k$ and associated $r$ for model 2. The carcass locations are shown as black circles, the blue polygon is the Etosha salt pan, the blue crosses are waterholes and the black lines are roads}
        \label{fig:zoomplots}
\end{figure}

The spatial term also contributed positively to the model, despite the extra parameters incurred (Table \ref{tab:1Dresult}); the BIC score decreased from 3084 for the univariate model (Model 1) to 2848 when the spatial term was included (Model 2). The practical consequences of its inclusion was clearly evidenced by tempering the `global' effect of roads and water which was implicit in the model that included the additional variables (Figure \ref{fig:mikeoutputs}a). In some cases the road and water effects diminished altogether where carcasses were not seen in the data. Crucially, this spatial term also better accommodates carcass locations which are not explained by only their proximity to water, distance to roads or average annual rainfall. Figure \ref{fig:zoomplots} shows that in Model 1, the water hole relationship dominates with a peak of intensity at each one. When the spatial term is added, the waterhole peak is suppressed at a number of waterholes and even increased at others. The peak in intensity is shifted to the north which is in keeping with the high number of carcasses observed there. The knot locations are similar to Model 3 but with fewer in the west and a higher proportion of smaller $r$ (Figure \ref{fig:zoomplots}b). Overall, the modelling shows that most, but not all, waterholes and some roads have high carcass intensity. Figure \ref{fig:mikeoutputs}b shows the top 5\% highest carcass intensity areas which form the highest risk areas in the park.

\begin{table}[!ht]
\centering
\caption{Table showing the results for the model based on one dimensional smoother-based relationships only (model 1) and the model with both one and two dimensional smoothers (model 2). For reference, model 3 is the model with only a two dimensional smooth (see Table \ref{tab:mrsearesult}).}
\vspace{0.5cm}
\begin{tabular}{c|c|c|r|c|c}
Model & Term  & $df$ & $\chi^2$ $p$-value  & Log-Likelihood & BIC\\\hline
1 & s(rainfall) & 3 & $p <$ 0.0001  &  -1505.4 & 3084.4\\
 & s(distRoads) & 3 & $p <$ 0.0001  &  & \\
 & Near water & 1 & $p <$ 0.0001  &  & \\\hline
2 & s(rainfall) & 3 &  $p <$ 0.0001  & -1221.3 & 2848.0\\
 & s(distRoads) & 3 & $p <$ 0.0001  & &\\
  & Near water & 1 & $p <$ 0.0001  & &\\
 & s(xcoord, ycoord) &  36 &  $p <$ 0.0001  & &\\\hline
3 & s(xcoord, ycoord) & 41 & $p <$ 0.0001 & -1301.6 & 2980.1\\\hline
\end{tabular}
\label{tab:1Dresult}
\end{table}


\clearpage

\section*{Discussion}\label{discussion}

Using SALSA2D for model selection provided better results and the ability to have a more realistic local/clustered intensity surface compared with the existing model averaging approach. There are clear numerical and practical benefits to SALSA2D-based model selection compared with a model averaging approach in this case and while the benefits of doing so might be less stark in cases where spatial patterns are more smooth, it needs to be possible to identify clusters and irregular patterns, such as those observed here, when they exist. 

Simply including proximity to water and roads in the model as part of this analysis did not reveal genuine patterns in all areas of the park, since not all roads/waterholes have been associated with carcasses. The addition of the spatial term with spatially adaptive knot selection was able to suppress/enhance the global relationships with the environmental covariates in particular areas. This resulted in the identification of some critical areas of the park which is important for effective park management, both in terms of disease outbreak, which after `unknown' was the largest category, and poaching, which although small in this data set (\textless{}1\% of carcasses), is increasing in the park. 

It is impossible to patrol such a large area at random and the areas of the park identified here (particularly those accessed by a subset of roads/waterholes) appear to require more monitoring efforts than others. Elephants are highly mobile and so early detection of carcasses, in particular anthrax related deaths, are important to identify spread of disease across the park \cite{Lind1994}.  The area of high intensity of carcasses to the south of the main pan matches well to the area of high anthrax risk identified by \citet{dougherty2022}.  This is also the area where the majority of anthrax or suspected anthrax cases were found in our database. Specifically, we show here that within this anthrax risk area the highest intensity of carcasses is near the waterholes.  

In the critical high carcass intensity area identified in the north-east of the park, the cause of death is less clear as the majority of carcasses were of unknown cause. However, it is interesting to note that the water sources in Etosha are a mix of boreholes and springs but the north east corner is mostly springs.  It is possible that in this region, there is higher water stress during drought which plays a role in mortality.

Whilst the density of elephant across the park is shown to be fairly constant \cite{Craig2020}, we have found that the density of carcasses is not.  Mortality is one of the key components in population dynamics models and the effects of spatial and temporal heterogeneity must be accounted for to have accurate predictive models for use in management and conservation \cite{sibly2009}. 

Therefore it is important to understand the changes in the spatial distribution of mortality across the park to have a better understanding of the population dynamics and drivers for population management.  For example, it is well known that surface water availability drives the distribution and abundance of elephants and that artificial manipulation of water availability is one of the tools available for the management of elephant populations \cite{chamaille2007}.  Closing waterholes is an option for managers to control elephant numbers should the numbers in Etosha continue to rise. One might base this decision on a number or factors including knowledge of which waterholes have a high density but low mortality. 

If the deaths are natural and, for instance, disease-related (e.g.~anthrax) then this provides valuable information about the prevalence and locale of disease in the park. Endemic anthrax occurs in Etosha annually \cite{Turner2013} and plays an important role in elephant population regulation/limitation. The monitoring of the prevalence of anthrax in elephant is important, because it advances our knowledge of a top down factor limiting a mega-herbivore. 

Even though the poaching of elephants in ENP is low (20 deaths reported to MIKE in 2018 and none poached), the general trend in more recent years is increasing (\textit{pers. comm.} Etosha Ecological Institute). As the number of poached elephants increases it is very useful knowledge to have a baseline distribution of natural deaths. It is also very important in light of the mass death events seen in Botswana in 2020 and 2021 \cite{Karombo2021}. Having a sense of ``normal'' places of natural death may provide insights should such events ever occur in Etosha.

\citet{critchlow2017} developed a method for improving the efficiency of ranger patrols using ranger collected monitoring data.  Ranger patrols are not just important for law enforcement but also the conservation of key species. With limited resource available for patrols, the key is to ensure that the patrol effort is efficient with respect to the activity one wishes to combat. The starting point for the method presented by \citet{critchlow2017} is a least one geographical map of illegal activity occurrence. However, the activity does not need to be an illegal one and in this case the activity of interest could be risk of disease outbreak. Along with a map of existing ranger effort, the carcass intensity maps presented here could be used to assess and improve the existing ranger effort in the park without the need for increased resources. 

Furthermore, should poaching increase in Etosha, then the methods presented here can provide necessary information about the prevalence, locale and patterns of these deaths. Additionally, should poaching occur in regions not common to find natural deaths, then increased/targeted mitigation measures can be efficiently actioned. Practically, understanding both the magnitude and spatial patterns of elephant deaths in ENP may assist in adapting patrol efforts in and around the park to track the anthrax disease and/or combat any poaching activities.

This extended CReSS approach using SALSA2D model selection presented in this paper is of immediate and practical value to a wide range of users of statistical modelling methods. SALSA2D is implemented inside the \texttt{MRSea} \texttt{R} package and it can automatically select knots based on two user-defined types of two-dimensional spline bases (Gaussian and exponential) and distance calculation (Euclidean or geodesic) based on a range of objective fitness criteria, chosen by the user. Notably, exclusion zones and non-Euclidean distances can be included to model more complex spatial regions \cite[as seen in][]{scott2014, scott2017} and adaptations have been made to allow for the fitting of Poisson PPMs using the downweighted Poisson regression method. By using presence only data in this paper, as opposed to the more traditional Binomial or Poisson GAMs, we have demonstrated the flexibility of this approach for a wide variety of settings.


\section*{Supplementary Material}\label{supplementary-material}

See Appendix S1 for information on residual calculation, pseudo-absence selection, expanded results of the methods comparison and details of rainfall interpolation.

\subsection*{Acknowledgements}
The authors would like to thank Dr. Richard Glennie for his assistance and patience with PPM queries and reading a draft of the paper. 

\subsection*{Notes on contributors}
\begin{itemize}
    \item LSH, MLM and CGW contributed to method development, analysis and paper writing
    \item GS, WK and PdP contributed to data collection and local information
\end{itemize}

\subsection*{Conflict of Interest}

There are no competing interests. 

\subsection*{Funding}
The authors declared that no grants were involved in supporting this work.


{\small\bibliographystyle{unsrtnat}
\bibliography{MIKEpaper_Ecosphere}}

\section*{Tables and Figures}

\bigskip

\end{document}



\title{Appendix S1: Supplementary Material for ``Automated surface feature selection using SALSA2D: An illustration using Elephant Mortality data in Etosha National Park." \\ Ecosphere}

\author{
\name{L.A.S Scott-Hayward\textsuperscript{a}\thanks{CONTACT L.A.S. Scott-Hayward. Email: lass@st-andrews.ac.uk}, M.L. Mackenzie\textsuperscript{a}, C.G. Walker\textsuperscript{b}, G. Shatumbu\textsuperscript{c}, W. Kilian\textsuperscript{c} and P. du Preez\textsuperscript{d}}
\affil{\textsuperscript{a}School of Mathematics and Statistics, University of St Andrews, KY16 9LZ, Fife, Scotland; \textsuperscript{b}Department of Engineering Science, University of Auckland, 70 Symonds Street, Auckland, New Zealand; \textsuperscript{c}Etosha Ecological Institute, PO Box 6, Okaukuejo via Outjo, Ministry of Environment, Forestry and Tourism, Namibia; \textsuperscript{d}African Wildlife Conservation Trust, PO box 97401, Windhoek, Namibia}
\affil{\textsuperscript{a}ORCiD: LSH (0000-0003-3402-533X), MLM (0000-0002-8505-6585)}
}

\maketitle 

\section{Finding the largest residual}

\begin{figure}[!htb]
\centering
\includegraphics[width=0.9\linewidth]{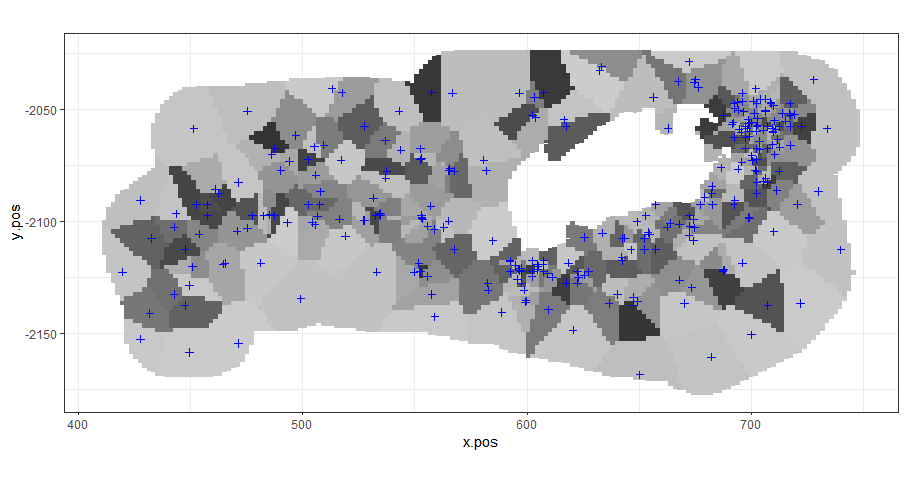}
\caption{Figure showing the knot locations (blue crosses) and the colour shows the nearest knot on the pseudo absence grid.}
\label{fig:blocks}
\end{figure}

\begin{itemize}
    \item Find nearest candidate knot location (of the legal knots remaining and ignoring the already selected knots) to each data point (both presence and pseudo-absence locations). Note that``nearest" is calculated based on whichever distance metric the model uses. Figure \ref{fig:blocks} shows the neighbourhood around each knot.
    \item Sum the observed counts within each knot region
    \item Make predictions to the pseudo absence grid and sum the estimated intensity within each knot region
    \item Calculate the absolute residual ($|(O-E)|$)
    \item Find the 10 knot regions with the largest score.  These become the candidates for the exchange/move step.
\end{itemize}

\newpage

\vspace{2cm}



\section{Pseudo-absence Selection}

\begin{itemize}
    \item Grid spacings trialled: 5, 4, 3, 2, 1.5, 1.25 and 1 km
    \item SALSA2D specification
    \begin{itemize}
        \item knot grid: all non-duplicated presence locations
        \item start knot number: 10, 20, 30, 40
        \item min knots and max knots equal to start knot number. 
        \item distance metric: Euclidean
        \item basis: Gaussian and exponential
    \end{itemize}
    \item Fit models for each specification of grid, start knots and basis
    \item Evaluate the log-likelihood
    \item Select the coarsest resolution after which, an increase in resolution makes little difference to the likelihood. 
\end{itemize}

\begin{figure}[!htb]
\centering
\includegraphics[width=0.9\linewidth]{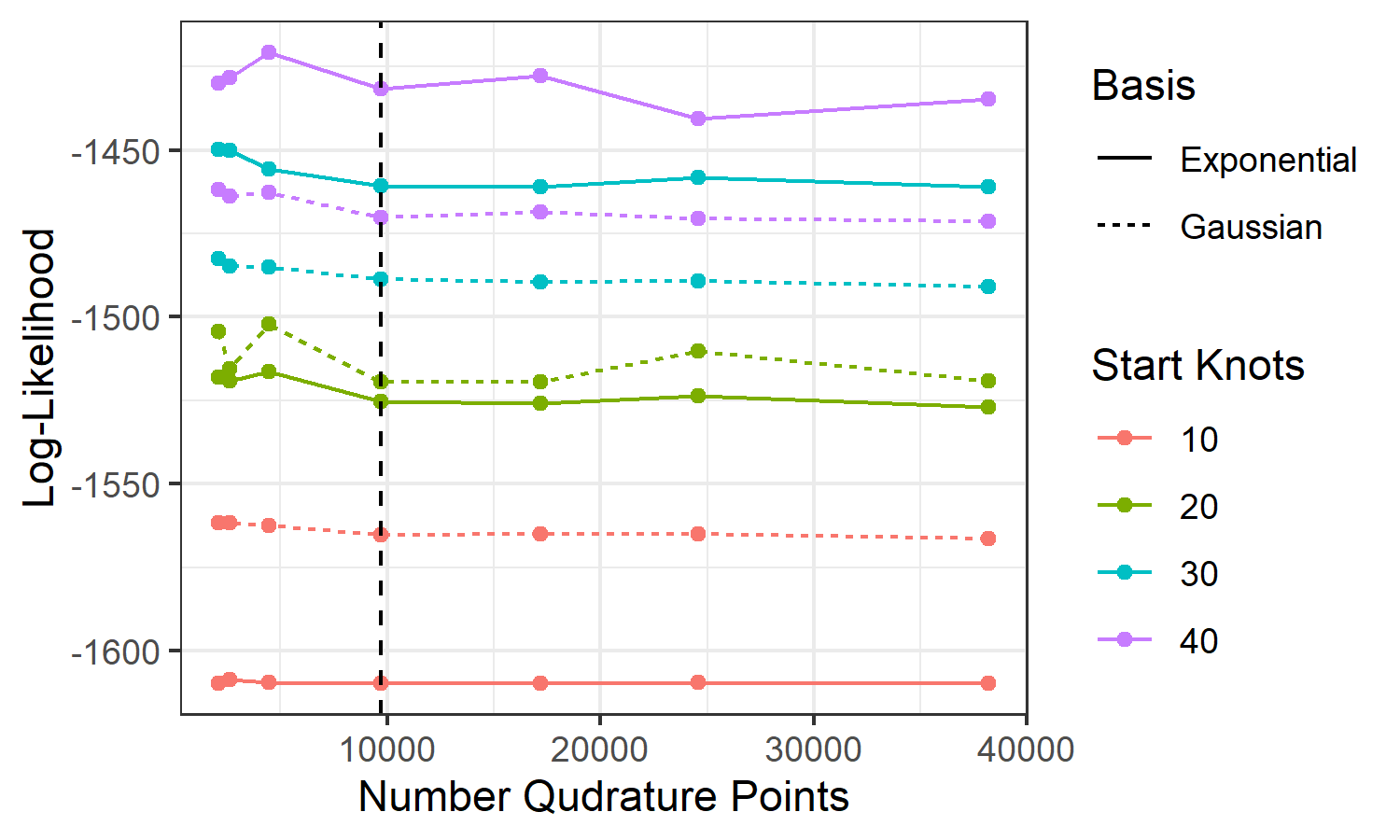}
\caption{Figure showing the convergence of the log-likelihood for for different spatial resolutions across multiple SALSA2D parameterisations.}
\label{fig:resolutionconvergence}
\end{figure}

Figure \ref{fig:resolutionconvergence} shows the log-likelihood scores for the different parameterisations. The vertical dashed line indicates the best grid resolution; 2km$^2$.

\section{Estimated Intensities}

\begin{figure}[!htb]
\centering
\includegraphics[width=0.9\linewidth]{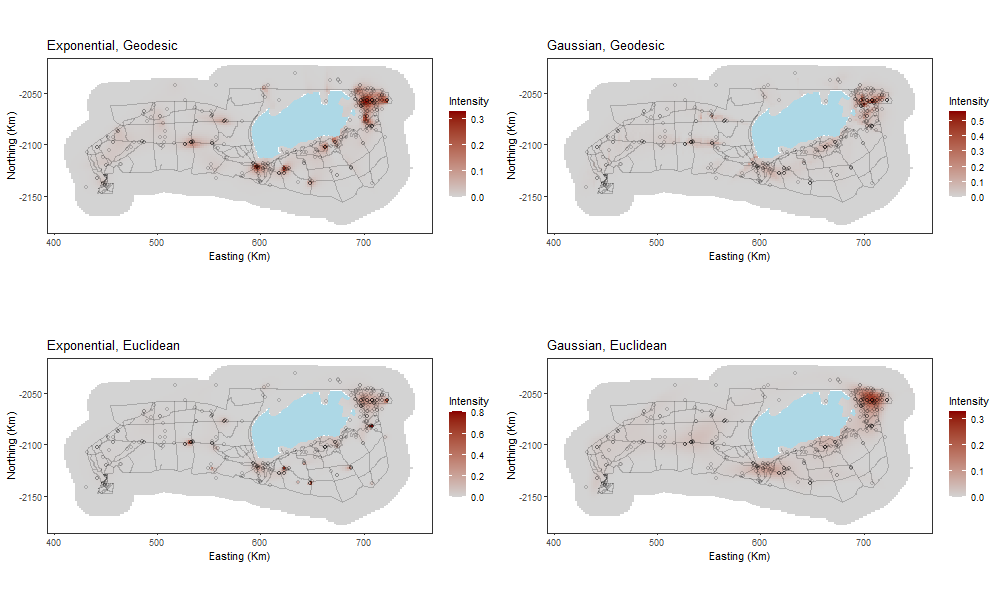}
\caption{Figure showing fitted intensity surfaces for the four SALSA2D models selected using log-likelihood.}
\label{fig:salsaloglik}
\end{figure}

The following figures show the best models from the two different methodological frameworks and the four different parameterisations. The best models are selected using BIC for SALSA2D and AIC$_c$ weights for model averaging. Since, the two frameworks are compared using the log-likelihood, Figure \ref{fig:salsaloglik} shows the best log-likelihood selected models. 


\begin{figure}[!htb]
\centering
\includegraphics[width=0.9\linewidth]{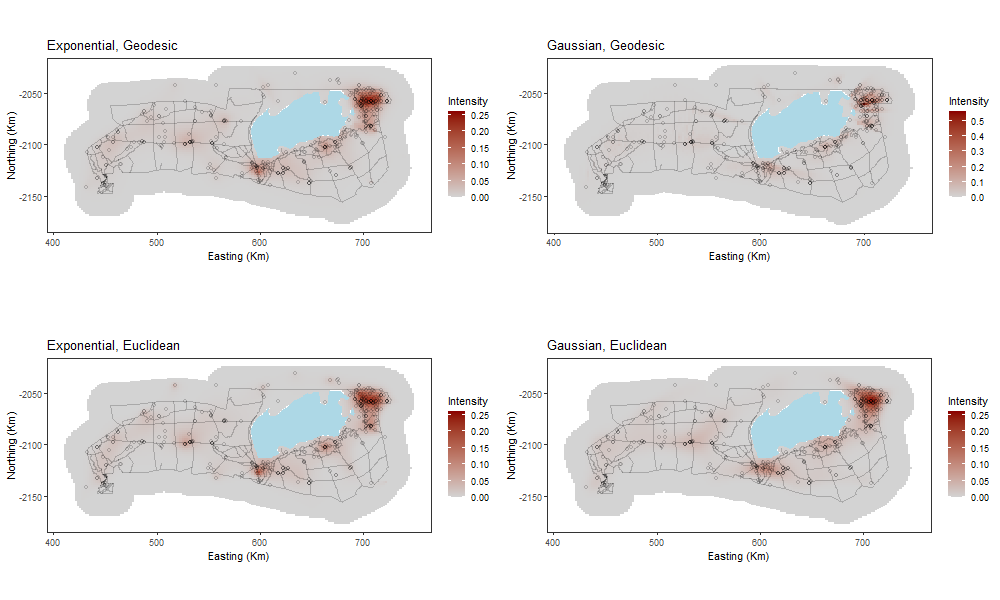}
\caption{Figure showing the best model averaged outputs from the four different parametrisations and selected using AIC$_c$ model weights.}
\label{fig:modelavgall}
\end{figure}

\clearpage

\subsection*{SALSA2D outputs}

\begin{table}[!htb]
\renewcommand{\arraystretch}{0.5}
\begin{tabular}{|l|l|r|r|r|r|r|}
\hline
Distance Type & Basis & \makecell[r]{Start \\Knots} & \makecell[l]{End \\Knots} & LogLik & BIC & \makecell[l]{Time \\(min)}\\
\hline
Euclidean & Exponential & 5 & 8 & -1528.5 & 3130.6 & 2.3\\
\hline
Euclidean & Exponential & 10 & 8 & -1509.1 & 3091.9 & 2.1\\
\hline
Euclidean & Exponential & 15 & 8 & -1517.0 & 3107.7 & 2.7\\
\hline
Euclidean & Exponential & 20 & 10 & -1504.6 & 3101.4 & 4.6\\
\hline
Euclidean & Exponential & 25 & 12 & -1486.3 & 3083.1 & 21.1\\
\hline
Euclidean & Exponential & 30 & 19 & -1445.0 & 3065.0 & 22.9\\
\hline
Euclidean & Exponential & 35 & 22 & -1438.7 & 3080.0 & 21.6\\
\hline
Euclidean & Exponential & 40 & 27 & -1407.7 & 3064.1 & 46.7\\
\hline
Euclidean & Exponential & 45 & 28 & -1376.8 & 3011.5 & 49.9\\
\hline
Euclidean & Exponential & 50 & 32 & -1374.5 & 3043.8 & 66.7\\
\hline
Euclidean & Exponential & 55 & 40 & -1318.0 & 3004.4 & 73.3\\
\hline
Euclidean & Exponential & 60 & 41 & -1301.6 & 2980.9 & 126.2\\
\hline
Euclidean & Gaussian & 5 & 7 & -1558.7 & 3181.9 & 1.4\\
\hline
Euclidean & Gaussian & 10 & 7 & -1572.4 & 3209.2 & 1.4\\
\hline
Euclidean & Gaussian & 15 & 8 & -1572.4 & 3218.4 & 2.1\\
\hline
Euclidean & Gaussian & 20 & 14 & -1533.1 & 3195.1 & 2.3\\
\hline
Euclidean & Gaussian & 25 & 15 & -1522.3 & 3182.8 & 18.4\\
\hline
Euclidean & Gaussian & 30 & 18 & -1515.9 & 3197.5 & 16.4\\
\hline
Euclidean & Gaussian & 35 & 28 & -1490.3 & 3238.6 & 37.6\\
\hline
Euclidean & Gaussian & 40 & 28 & -1489.0 & 3235.9 & 38.3\\
\hline
Euclidean & Gaussian & 45 & 34 & -1479.6 & 3272.5 & 31.6\\
\hline
Euclidean & Gaussian & 50 & 39 & -1471.3 & 3301.9 & 23.8\\
\hline
Euclidean & Gaussian & 55 & 47 & -1452.3 & 3337.4 & 38.1\\
\hline
Euclidean & Gaussian & 60 & 47 & -1451.6 & 3336.2 & 103.9\\
\hline
\end{tabular}
\caption{SALSA2D outputs for the Euclidean distance models\label{tab:outputseuc}}
\end{table}

\begin{table}[!htb]
\begin{tabular}{|l|l|l|r|r|r|r|r|}
\hline
  & Distance Type & Basis & \makecell[r]{Start \\Knots} & \makecell[l]{End \\Knots} & LogLik & BIC & \makecell[l]{Time \\(min)}\\
\hline
25 & Geodesic & Exponential & 5 & 9 & -1502.5 & 3087.8 & 2.2\\
\hline
26 & Geodesic & Exponential & 10 & 5 & -1539.3 & 3124.7 & 2.7\\
\hline
27 & Geodesic & Exponential & 15 & 9 & -1492.5 & 3067.9 & 4.1\\
\hline
28 & Geodesic & Exponential & 20 & 12 & -1481.8 & 3074.1 & 5.1\\
\hline
29 & Geodesic & Exponential & 25 & 13 & -1472.7 & 3065.2 & 21.5\\
\hline
30 & Geodesic & Exponential & 30 & 15 & -1465.8 & 3069.8 & 18.6\\
\hline
31 & Geodesic & Exponential & 35 & 19 & -1448.3 & 3071.6 & 40.7\\
\hline
32 & Geodesic & Exponential & 40 & 25 & -1415.3 & 3060.9 & 13.5\\
\hline
33 & Geodesic & Exponential & 45 & 21 & -1427.4 & 3048.2 & 45.8\\
\hline
34 & Geodesic & Exponential & 50 & 32 & -1369.7 & 3034.2 & 62.7\\
\hline
35 & Geodesic & Exponential & 55 & 24 & -1398.1 & 3017.3 & 90.7\\
\hline
36 & Geodesic & Exponential & 60 & 28 & -1377.6 & 3013.1 & 103.7\\
\hline
37 & Geodesic & Gaussian & 5 & 8 & -1551.0 & 3175.6 & 1.2\\
\hline
38 & Geodesic & Gaussian & 10 & 6 & -1562.4 & 3180.1 & 0.9\\
\hline
39 & Geodesic & Gaussian & 15 & 7 & -1553.0 & 3170.6 & 0.8\\
\hline
40 & Geodesic & Gaussian & 20 & 11 & -1510.0 & 3121.4 & 18.5\\
\hline
41 & Geodesic & Gaussian & 25 & 14 & -1500.4 & 3129.8 & 19.8\\
\hline
42 & Geodesic & Gaussian & 30 & 15 & -1481.7 & 3101.6 & 13.9\\
\hline
43 & Geodesic & Gaussian & 35 & 21 & -1472.5 & 3138.4 & 33.7\\
\hline
44 & Geodesic & Gaussian & 40 & 25 & -1450.5 & 3131.3 & 8.7\\
\hline
45 & Geodesic & Gaussian & 45 & 26 & -1433.8 & 3107.0 & 46.9\\
\hline
46 & Geodesic & Gaussian & 50 & 31 & -1418.0 & 3121.7 & 40.9\\
\hline
47 & Geodesic & Gaussian & 55 & 32 & -1408.3 & 3111.3 & 76.8\\
\hline
48 & Geodesic & Gaussian & 60 & 33 & -1411.8 & 3127.5 & 26.6\\
\hline
\end{tabular}
\caption{SALSA2D outputs for the Geodesic models}
\label{tab:outputsgeo}
\end{table}

\begin{table}[!htb]
\begin{tabular}{|l|l|r|r|r|r|r|}
\hline
Distance Type & Basis & \makecell[r]{Start \\Knots} & \makecell[l]{End \\Knots} & LogLik & BIC & \makecell[l]{Time \\(min)}\\
\hline
Euclidean & Exponential & 60 & 41 & -1301.6 & 2980.9 & 126.2\\
\hline
Euclidean & Exponential & 55 & 40 & -1318.0 & 3004.4 & 73.3\\
\hline
Geodesic & Exponential & 50 & 32 & -1369.7 & 3034.2 & 62.7\\
\hline
Euclidean & Exponential & 50 & 32 & -1374.5 & 3043.8 & 66.7\\
\hline
Euclidean & Exponential & 45 & 28 & -1376.8 & 3011.5 & 49.9\\
\hline
Geodesic & Exponential & 60 & 28 & -1377.6 & 3013.1 & 103.7\\
\hline
Geodesic & Exponential & 55 & 24 & -1398.1 & 3017.3 & 90.7\\
\hline
Euclidean & Exponential & 40 & 27 & -1407.7 & 3064.1 & 46.7\\
\hline
Geodesic & Gaussian & 55 & 32 & -1408.3 & 3111.3 & 76.8\\
\hline
Geodesic & Gaussian & 60 & 33 & -1411.8 & 3127.5 & 26.6\\
\hline
\end{tabular}
\caption{Top 10 log-likelihood selected SALSA2D models}
\end{table}

\begin{table}[!htb]
    \centering
\begin{tabular}{|l|l|r|r|r|r|r|}
\hline
Distance Type & Basis & \makecell[r]{Start \\Knots} & \makecell[l]{End \\Knots} & LogLik & BIC & \makecell[l]{Time \\(min)}\\
\hline
Euclidean & Exponential & 60 & 41 & -1301.6 & 2980.9 & 126.2\\
\hline
Euclidean & Exponential & 55 & 40 & -1318.0 & 3004.4 & 73.3\\
\hline
Euclidean & Exponential & 45 & 28 & -1376.8 & 3011.5 & 49.9\\
\hline
Geodesic & Exponential & 60 & 28 & -1377.6 & 3013.1 & 103.7\\
\hline
Geodesic & Exponential & 55 & 24 & -1398.1 & 3017.3 & 90.7\\
\hline
Geodesic & Exponential & 50 & 32 & -1369.7 & 3034.2 & 62.7\\
\hline
Euclidean & Exponential & 50 & 32 & -1374.5 & 3043.8 & 66.7\\
\hline
Geodesic & Exponential & 45 & 21 & -1427.4 & 3048.2 & 45.8\\
\hline
Geodesic & Exponential & 40 & 25 & -1415.3 & 3060.9 & 13.5\\
\hline
Euclidean & Exponential & 40 & 27 & -1407.7 & 3064.1 & 46.7\\
\hline
\end{tabular}
\caption{Top 10 BIC selected SALSA2D models.}
    \label{tab:my_label}
\end{table}

\clearpage

\subsection*{Model Averaged Selected Models}

\begin{itemize}
    \item $k$ = number of knots
    \item $r$ = effective radius sequence number
    \item $w$ = AIC$_c$ model averaging weight
\end{itemize}

\begin{verbatim}
$expeuc
   k r w
1 50 1 1

$expgeo
    k  r           w
1  45  3 0.003343363
2  50  1 0.131329653
3  50  3 0.070500968
4  50  4 0.070848736
5  50  5 0.058365475
6  50  6 0.055646771
7  50  7 0.055112831
8  50  8 0.055009183
9  50  9 0.054987913
10 50 10 0.054925003
11 60  1 0.389930105

$gauseuc
   k r           w
1 55 2 0.006596805
2 55 3 0.052268277
3 55 4 0.062721223
4 55 5 0.025132387
5 60 2 0.107459248
6 60 3 0.566550328
7 60 4 0.159010615
8 60 5 0.020261116

$gausgeo
   k r          w
1 60 6 0.98712668
2 60 7 0.01287332
\end{verbatim}





\section{Full Analysis Covariate Information}

\subsection*{Rainfall calculation} 

Fit a high dimensional smooth term to 156 locations of annual rainfall from 1999 to 2015 (2016/17 unavailable at the time of modelling) to interpolate values for the presence locations and pseudo-absence grid.

\begin{verbatim}
require(mgcv)
fit<-gam(meanrain ~ s(x.pos, y.pos,fx = TRUE, k=150), data=rainfall2)
analysisdat$meanrain<-predict(object = fit, 
                    newdata = data.frame(x.pos = analysisdat$x.pos, 
                                        y.pos = analysisdat$y.pos))
\end{verbatim}

\begin{figure}[!htb]
\centering
\includegraphics[width=0.9\linewidth]{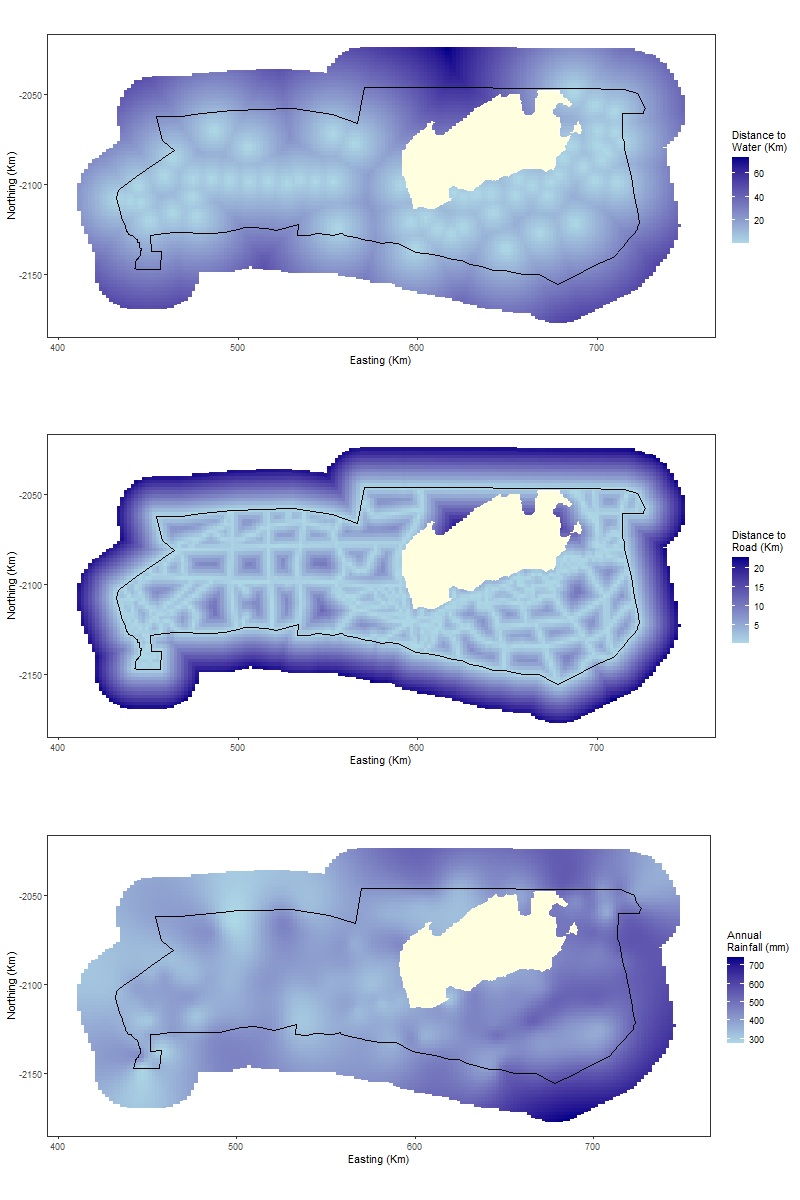}
\caption{Figure showing the covariate data in the study region. Distance to water (top), distance to roads (centre) and interpolated annual rainfall (bottom).}
\label{fig:covariates}
\end{figure}

The results of the interpolation model are shown in Figure \ref{fig:covariates}

